\documentclass[aps,twocolumn,showpacs,preprintnumbers,amsmath,amssymb,prb,superscriptaddress]{revtex4-1}
\usepackage{graphicx}
\usepackage{bm}
\usepackage{color, colortbl}
\usepackage{amsmath}
\usepackage{amsfonts}
\usepackage{amssymb}

\newcommand{\bea}{\begin{eqnarray}}
\newcommand{\enea}{\end{eqnarray}}
\newcommand{\beq}{\begin{equation}}
\newcommand{\eneq}{\end{equation}}

\definecolor{Gray}{gray}{0.9}
\newcolumntype{d}{>{\columncolor{Gray}}c}
\newcolumntype{m}{>{\columncolor{Gray}}l}
\newcolumntype{s}{>{\columncolor{Gray}}r}

\newcommand{\TBKT}{T_\mathrm{BKT}}

\begin{document}

\title{Incipient Berezinskii-Kosterlitz-Thouless transition in two-dimensional coplanar Josephson junctions}

\author{D. Massarotti}
\email{dmassarotti@na.infn.it}
\affiliation{Dipartimento di Ingegneria Industriale e dell'Informazione, Seconda Universit\`a  di Napoli, I-81031 Aversa (CE), Italy}
\affiliation{CNR-SPIN UOS Napoli, Monte S. Angelo-Via Cintia, I-80126, Napoli, Italy}

\author{B. Jouault} 
\affiliation{Laboratoire Charles Coulomb (L2C), UMR 5221 CNRS-Universit\'e de Montpellier, F-34095, Montpellier, France}

\author{V. Rouco} 
\affiliation{Dipartimento di Fisica "E. Pancini", Universit\`{a} di Napoli Federico II, Monte S. Angelo-Via Cintia, I-80126 Napoli, Italy}

\author{S. Charpentier}
\affiliation{Department of Microtechnology and Nanoscience, Chalmers University of Technology, SE-412 96 G\"oteborg, Sweden}

\author{T. Bauch}
\affiliation{Department of Microtechnology and Nanoscience, Chalmers University of Technology, SE-412 96 G\"oteborg, Sweden}

\author{A. Michon}
\affiliation{Centre de Recherche sur l'H\'et\'ero-Epitaxie et ses Applications (CRHEA), CNRS, rue Bernard Gr\'egory, 06560 Valbonne, France} 

\author{A. De Candia}
\affiliation{CNR-SPIN UOS Napoli, Monte S. Angelo-Via Cintia, I-80126, Napoli, Italy}
\affiliation{Dipartimento di Fisica "E. Pancini", Universit\`{a} di Napoli Federico II, Monte S. Angelo-Via Cintia, I-80126 Napoli, Italy}

\author{P. Lucignano}
\affiliation{CNR-SPIN UOS Napoli, Monte S. Angelo-Via Cintia, I-80126, Napoli, Italy}
\affiliation{Dipartimento di Fisica "E. Pancini", Universit\`{a} di Napoli Federico II, Monte S. Angelo-Via Cintia, I-80126 Napoli, Italy}

\author{F. Lombardi}
\affiliation{Department of Microtechnology and Nanoscience, Chalmers University of Technology, SE-412 96 G\"oteborg, Sweden}

\author{F. Tafuri}
\affiliation{Dipartimento di Ingegneria Industriale e dell'Informazione, Seconda Universit\`a  di Napoli, I-81031 Aversa (CE), Italy}
\affiliation{CNR-SPIN UOS Napoli, Monte S. Angelo-Via Cintia, I-80126, Napoli, Italy}

\author{A. Tagliacozzo}
\affiliation{CNR-SPIN UOS Napoli, Monte S. Angelo-Via Cintia, I-80126, Napoli, Italy}
\affiliation{Dipartimento di Fisica "E. Pancini", Universit\`{a} di Napoli Federico II, Monte S. Angelo-Via Cintia, I-80126 Napoli, Italy}
\affiliation{INFN, Laboratori Nazionali di Frascati, Via E.Fermi, Frascati, Italy}


\begin{abstract}

Superconducting hybrid junctions are revealing a variety of novel effects. Some of them are due to the special layout of these devices, which often use a coplanar configuration with relatively large barrier channels and the possibility of hosting Pearl vortices.  A Josephson junction with a quasi ideal two-dimensional barrier has been realized by growing graphene on SiC with Al electrodes. Chemical Vapor Deposition offers centimeter size monolayer areas  where it is possible to realize a comparative analysis of different devices with nominally the same barrier.  In samples with a graphene gap below 400 nm, we have found evidence of  Josephson coherence in presence of an incipient  Berezinskii-Kosterlitz-Thouless transition.  When the magnetic field is cycled, a remarkable  hysteretic collapse and revival of the Josephson supercurrent occurs. Similar hysteresis are found in granular systems and are usually justified within the Bean Critical State model (CSM). We show that the CSM, with appropriate account for the low dimensional geometry, can partly explain the odd features measured in these junctions. 

\end{abstract}


\maketitle
  
\section{Introduction}


Superconducting hybrid junctions
can now be obtained combining superconductors with functional barriers as semiconductors \cite{doh}, graphene\cite{revue09,Beenakker_Graphene}, topological insulators \cite{Qi-Zhang,Hasan_Kane}, ferromagnets\cite{Bergeret_Volkov_Efetov,Buzdin} made superconducting by the proximity effect. 
These devices often use a coplanar layout with almost two-dimensional (2D) flakes  or nanowires as barriers.  \cite{doh,maj3,kurter,Galletti:2014}  
The geometry of the device and the nature of the interfaces may favor the appearance of novel exotic  effects ranging from Majorana fermions \cite{Alicea:2012,Beenakker_MF} to topological superconductivity \cite{Hasan_Kane}.
In this paper we will show how the emerging class of 2D extended barriers can promote Josephson coherence in presence of an incipient  Berezinskii-Kosterlitz-Thouless \cite{Berezinski1972JETP,Kosterlitz1973JPC} transition, which is indeed, supposed to be a peculiarity of 2D superconducting systems. 
In this case a 2D barrier is made superconducting by proximity effect. To this aim we have realized graphene Josephson junctions (GJJs), obtained using a graphene barrier deposited on SiC by Chemical Vapor Deposition (CVD) rather than the standard technique of an exfoliated flake\cite{noi16}. 
Graphene on SiC  is highly homogeneous at the centimeter scale. 
Thus relatively simple lithography processes allow to obtain thousands of devices on the same wafer.\cite{Kedzierski2008IEEE}
This specific type of graphene growth guarantees to have devices with the desired geometry, 
{\it i.e.} a 2D barrier of suitable lateral dimensions to host 2D vortex structures.  Large graphene samples deposited as a thin film offers enough room to host  extremely spread out 2D vortex entities.
Vortex pinning is expected to occur because of interface impurities between the graphene sheet and the thin superconducting electrode.

In this paper we show how the Josephson coherence is observed along with very unusual features like the persistence of a small residual resistance and, even more surprising, the appearance of a hysteresis in magnetic field, manifesting in collapses and revivals of the supercurrent depending on the direction of the magnetic sweeping (see Fig.~\ref{fig_butterflies}). This points to Josephson conduction in presence of an incipient  Berezinskii-Kosterlitz-Thouless (BKT) transition at subKelvin temperature. 

Our results are a  fundamental step towards scalability for GJJs, and for our goals it offers the additional advantage of comparing junctions fabricated on exactly the same barrier. This can be extremely useful to have further insights on the electrodynamic response and the nature of dissipation of GJJs, which need to be considered in the actual rush of more performing devices
%
in the ballistic limit, to exploit the unique properties of Andreev reflection in graphene.\cite{Beenaker2006PRL,TitovOssipovPRB2007,Black-ShafferPRB2008,DirksNatphys2011,BenShalom1504.03286,MizunoNCOM2013,CaladoNNANO2015}

In Section II we report  on the preparation of the sample. 
A consistent interpretation of the phenomenology described in Sections III--V requires frames which  go beyond the usual Josephson S/N/S paradigm. The superconducting proximity  is quite unique in  our samples.  The Josephson supercurrent  is always accompanied by a small resistance, but it drops to a very low value very fast, as soon as a very weak orthogonal magnetic field $H$ is applied, apparently without  loosing Josephson phase coherence.
Meanwhile, the resistance increases from a few Ohms  to values of the order of
$400$~$\Omega$. The situation does not change if the field is further increased, but, as soon as we invert the sweeping direction at any magnetic field value below a threshold field,
the Josephson supercurrent has an unexpected revival, in presence of a remanent magnetization.  Simultaneously, the resistance drops again. 
All these observations are consistent with a special regime of the proximity effect,  expected to occur only in two dimensions  and in presence of pinning of  vortices. 


Similar hysteresis in the magnetization or in the critical current of granular systems is found in type II superconductors and is usually justified within the Bean Critical State model (CSM).\cite{Bean1962PRL}
The CSM can provide a qualitative interpretation of  the magnetization process, which takes place in  the Al/Ti  islands covering the graphene, and of the measured hysteresis.  
Support to this interpretation comes from the fact that  metal decorated graphene sheets have been fabricated\cite{Kessler2010PRL} and their superconducting phase transition has been classified as of the  BKT type.\cite{Berezinski1972JETP,Kosterlitz1973JPC}
Cooling down the devices below the critical temperature of the electrodes ($T< T_c^{Al} \simeq 1.1 $ K)  in zero field, apparently leads to incipient  superconductivity in the graphene layer. 

The plots of the resistance  {\it vs.} temperature presented in Section III support a qualitative interpretation of the measured phenomena within the  BKT theory (see  Fig.~\ref{fig_AL_BKT}) at low temperatures. In this regime vortex-antivortex ($v$-$\bar v$) pair unbinding  and flux-flow resistance across the junction are dominant phenomena.
We have measured our samples down to $T=280$ mK which is a much higher temperature than  the expected BKT critical temperature, $ \TBKT$. Features of the BKT  incipient transition survive in the crossover to a paraconductivity regime at $T_{BKT} < T< T_c^{Al} $  . The Josephson supercurrent modulated by the magnetic field is accompanied by the flow of vortices, which induces a finite slope in the supercurrent branch.
%
%
An analysis of the flux flow resistance points to large Pearl vortices, pinned to the impurities in the CSM phase. Magnetic screening  is very weak and correlations at intermediate distances induced by the long range repulsive interaction can be strong,  particularly when the applied orthogonal magnetic field is very small. All of this is discussed in Sections IV and V.

 In Section VI the Fraunhofer pattern of the Josephson conduction is discussed. By ignoring details of the microscopic vortex structure and dynamics,  a macroscopic approach to the diamagnetic  screening currents, based on the solution of the London equation in a quasi 2D contact, can reasonably account for the measured pattern if flux focusing effects  are assumed in the planar thin-film  weak link\cite{rosenthal}. Sections VII and Section VIII contain a detailed summary and the conclusions, respectively.   
 
\section{Sample preparation}

The graphene layer was grown by CVD 
on the silicon face of a SiC substrate.
The semi-insulating SiC substrate has an extremely small, negligible conductivity.
The quality of the monolayer structure was carefully investigated by Raman and 
Angle Resolved Photo Emission Spectroscopy\cite{noi16} (see Refs.~\onlinecite{Michon2010APL,Jabakhanji2014PRBa} for details).
Before lithography, the mobility at room temperature is
$\mu \simeq 500$ cm$^2$/V$\cdot$s. The graphene  is $p$-doped, 
$p \simeq 5 \times 10^{12}$ cm$^{-2}$.
This $p$-doping is induced by the hydrogenation of interface between graphene and SiC.\cite{Jabakhanji2014PRBa}

\begin{figure}
\includegraphics[width=0.95 \linewidth]{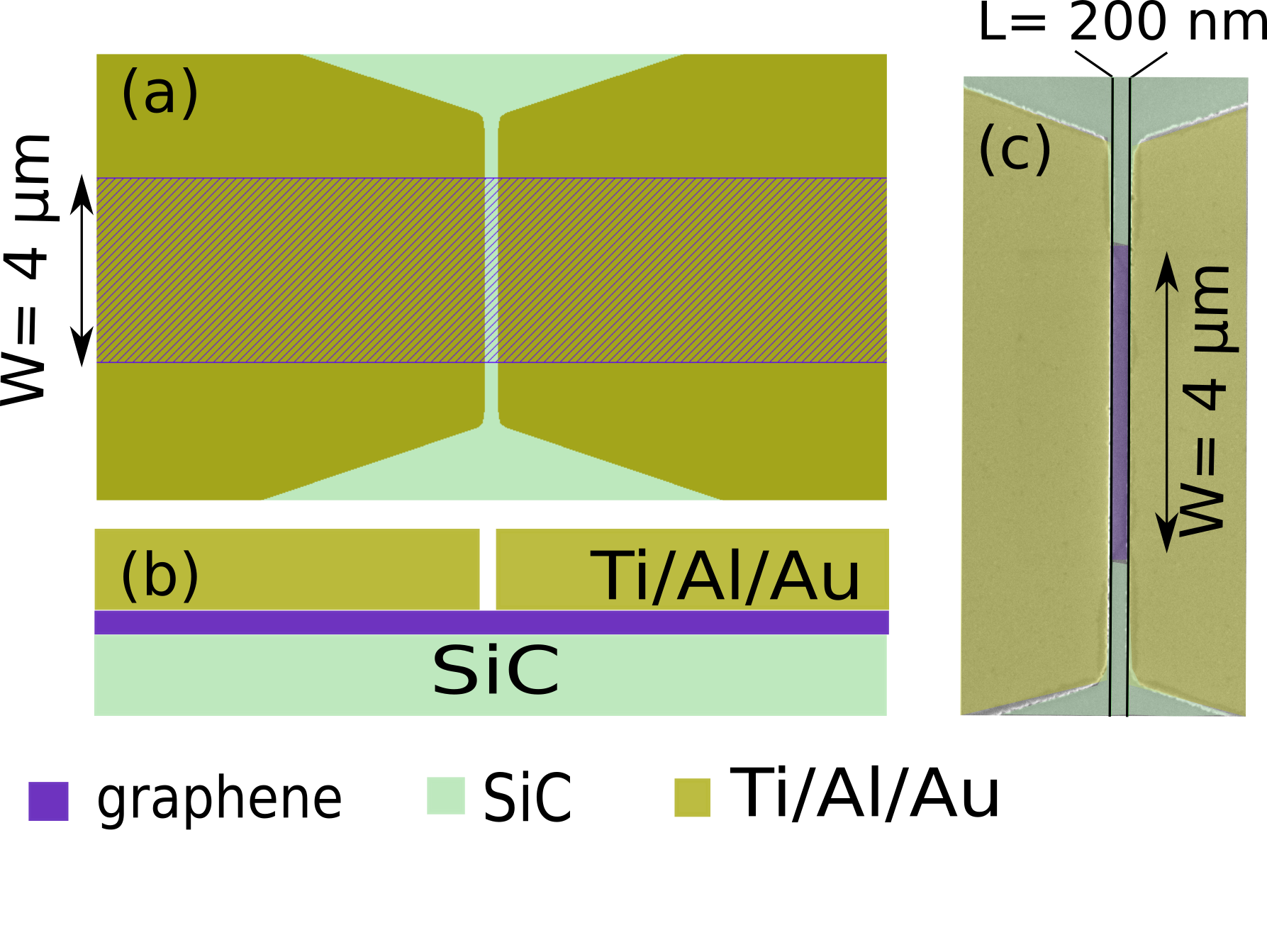}
\caption{{ Josephson devices with large 2D contact barriers.}{ (a)} Large sketch of the planar devices. { (b)} Sketch of the graphene junction. { (c)} Scanning-electron micrograph of a $L\approx 200$ nm short junction with artificial colors.}
\label{fig:sketch}
\end{figure} 


The planar Josephson devices are fabricated by conventional $e$-beam lithography. 
The graphene junctions are patterned by oxygen plasma. 
Then, the contacts are deposited (see Fig. \ref{fig:sketch}). 
They consist of a 5 nm interfacial layer of titanium, which ensures good electrical contact to graphene,
a 80 nm thick aluminium layer and a 3 nm thick gold layer.  
 It has been reported that Al on top of graphene could induce a strong $n-$doping\cite{Giovannetti2008PRL}. As the pristine graphene is initially $p$-doped, these Al/graphene junctions could be suitable  for fabricating $n$-$p$-$n$ junctions, with appropriate protocols which include gates.  In our devices, the gap $L$ between the electrodes remains relatively large, ranging from 200 nm to 600 nm. The width of the junctions is fixed at $W = 4~\mu$m. Charge transport is most likely diffusive. Thus Fabry-Perot resonances and Klein tunnelling are not expected to be present. 

 These junctions have a major difference with graphene junctions obtained by
the exfoliation technique. While the geometrical junction area is  ${\cal{A}}_j = W \times L$, here  the remaining graphene area
below the superconducting Al contacts is massive. On purpose, we left a graphene area
$\simeq 10^3$ $\mu$m$^2$ under each of the two Al contacts.
A part of these areas is visible in Fig.~\ref{fig:sketch}a.
For the measurements, the samples are thermally anchored to the cold stage of a 
$^3$He cryostat equipped with EMI filters at room temperature, RC filters at the 1-K pot stage and copper powder filters at the sample stage \cite{luigi,davide3}. Current-voltage (I-V) characteristics of different junctions have been measured as a function of temperature and magnetic field. Measurements of resistance as a function of temperature, $R(T)$, and magnetic field, $R(H)$, have been performed with standard low frequency lock-in techniques using low excitation currents, in the range 5--10 nA.

\begin{figure}
\begin{center}
\includegraphics[width=0.95 \linewidth]{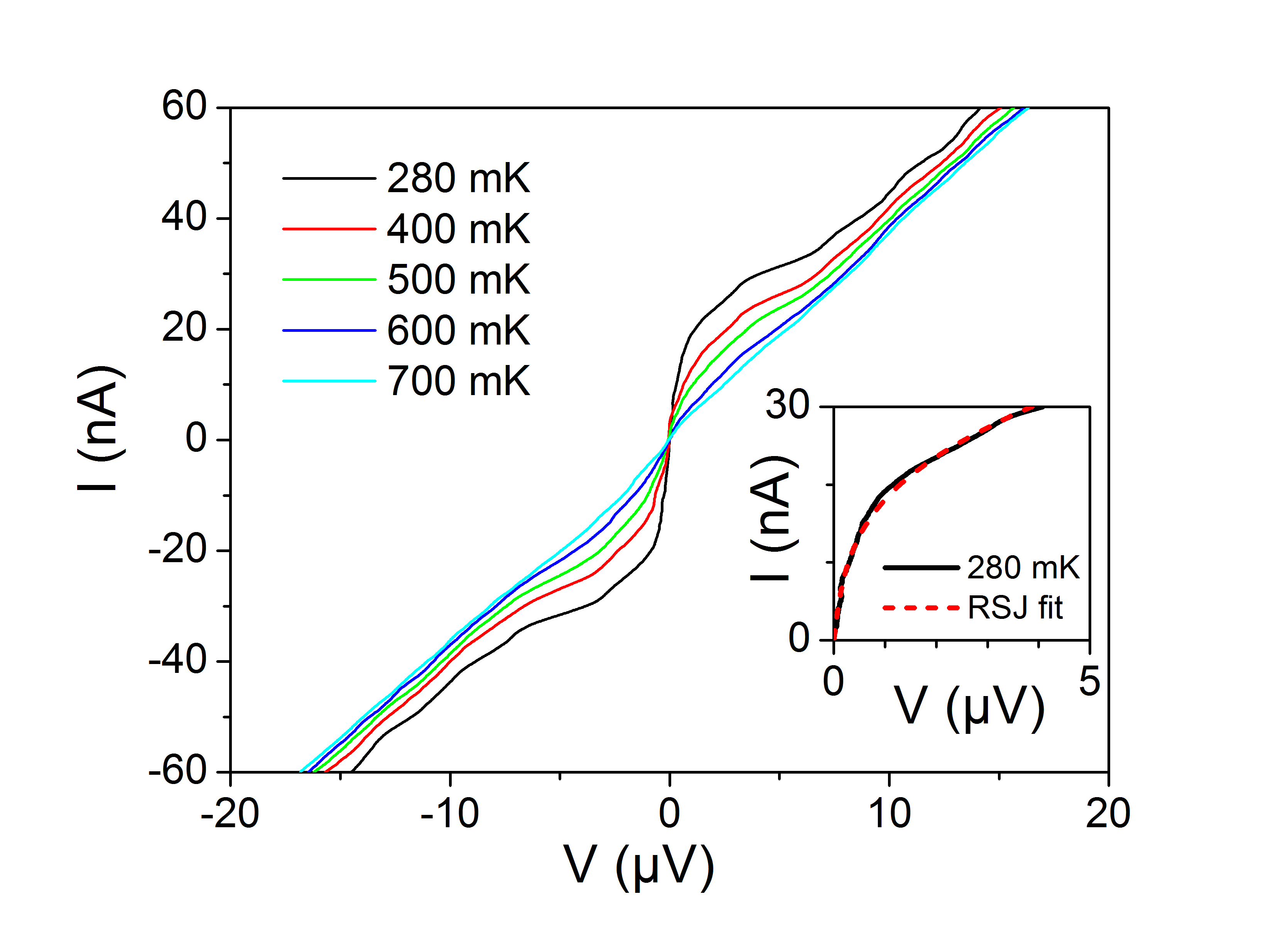}
\caption{{ I-V characteristics measured at various temperatures for the J200-4 junction.} The inset shows the RSJ fit ({\it red dashed line}) of the I-V curve measured at 280 mK ({\it black full line}).}
\label{fig_VI}
\end{center}
\end{figure}

Table \ref{table1} collects measured and fitted parameters of some of the studied junctions.
For the shortest junction (J200-4, $L = 200$ nm), the I-V curves are reported at various temperatures in Fig.~\ref{fig_VI}. We measure a finite slope in the I-V characteristics close to zero voltage, even at the lowest temperatures. The linear part is followed by a bending which is characteristic of the resistively shunted junction (RSJ) model. The critical current $I_c$ as a function of an externally applied magnetic field is estimated through the RSJ model (see inset of Fig.~\ref{fig_VI}), suitably used for an overdamped Josephson junction \cite{barone} in series with a resistance. We attribute the finite slope in the supercurrent branch to the presence of fluctuating broken vortex-antivortex pairs which cannot be described appropriately within the RSJ framework or more refined arguments based on phase diffusion \cite{luigi,davide3,barone,kautz} (see Sections \ref{sec_hysteresis} and \ref{collapse}). Indeed, our RSJ fit points to an effective temperature which is larger than the base temperature.  

\label{sec_samples}
 
 \section{Temperature dependence of the Resistance at zero $H$ field}

In Fig.~\ref{fig_AL_BKT}b the $R(T)$ curves in zero field cooling (ZFC) are plotted for the 3 junctions, which appear to be on the verge of a BKT transition at low temperatures (J200-3, J200-4  and J300-3). 
There is first a partial drop in the resistance which occurs just above $1$ K, that can be attributed to the transition to the superconducting state of the Al/Ti contacts. Below 1~K we identify two different regimes which we attribute to  Aslamazov--Larkin (AL) paraconductivity and incipient BKT transition. The two regimes are described here below. 

  \subsection{AL paraconductivity regime}
	
\begin{figure}
\includegraphics[width=0.95 \linewidth]{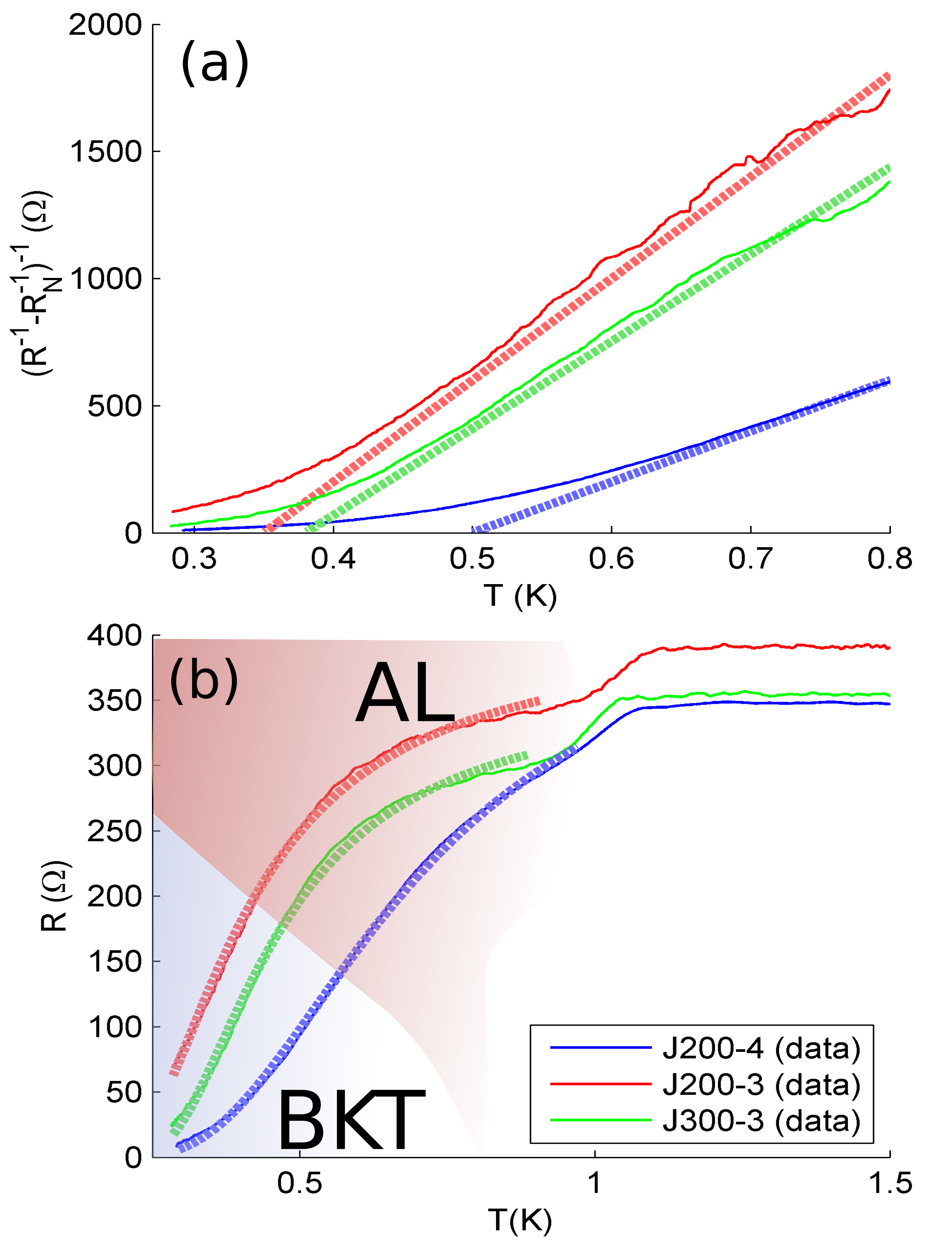}
\caption{{$R(T)$ curves of 3 junctions (J200-3, J200-4 and J300-3). (a)} The resistances are rescaled as $(R(T)^{-1}-R_N^{-1})^{-1}$  ({\it thin solid curves}). According to the AL theory, see  Eq.~\ref{ro}, interception of the linear fits ({\it thick solid lines}) with the $x$-axis gives the mean field pairing temperature $T_{c0}$.  { (b)} Fit of the resistances $R(T)$ of the 3 junctions given above, using  Eq.~\ref{halpNels},  with parameters from Table~\ref{table1} ({\it thick broken lines}). The domain of validity of the AL model is shaded in red. At lower temperature, the assumed domain of validity of 
the BKT theory is shaded in light blue.}
\label{fig_AL_BKT}
\end{figure}

 The high temperature partial drop in the resistance occurring  as 
 a  broad transition between 1 K and 0.5 K  is  consistent with a manifestation of  paraconductivity  in the graphene layer.
Thermal fluctuations in Cooper pair formation enhanced by proximity allow to define
a mean field pairing temperature $ T_{c0} $,
below which the amplitude of the superconducting condensation of pairs is expected to be finite.  
The drop  at zero magnetic field ($ T > T_{c0}$) can be fitted   by  the conductivity change $\delta \sigma (T)  \approx \ln^{-1}(T/T_{c0})$  typical of Aslamazov--Larkin (AL) fluctuation-enhanced conductivity in two dimensions.\cite{Aslamasov1968PLA} 
We replot  the data as $(R(T)^{-1}-R_N^{-1})^{-1}$
[{\sl thin solid lines  in Fig.~\ref{fig_AL_BKT}a)}].
According to the AL theory, linear fits [{\sl thick solid lines  in Fig.~\ref{fig_AL_BKT}a)}] give the value for the hypothetical mean field pairing temperature $T_{c0}$ at the interception with the $x$-axis\cite{Fiory1983PRB}:
\beq
(R(T)^{-1}-R_N^{-1})^{-1} = R_0  \left(  T - T_{c0} \right)/ T_{c0}.
\label{ro}
\eneq
Here  $R_0$ is a fitting parameter
predicted to be of the order of $16 \hbar /e^2$
while 	
$R_N$ is the normal resistance given by the slope of the I-V characteristics at large voltages. 
The parameters $R_0$ and $T_{c0}$ of the various measured junctions are reported in Table \ref{table1}.  
The mean-field pairing temperatures $T_{c0}$ which we obtain  in most of the measured samples fall within a small range of temperature $T= 0.23-0.5$ K.

\begin{table}
\caption{\label{table1} Main parameters for the investigated devices:
name, length $L$, normal resistance $R_N$, 
critical current $I_c$ at $T$= 280 mK,
mean field resistance $R_0$,  
mean field critical temperature $T_{c0}$,
BKT temperature, 
dimensionless parameter $b$ of Eq. (\ref{halpNels}). 
}
\begin{tabular}{|c| c d c d c d c| }
\hline
   name   & $L$&$R_N$&$I_c$&$R_0$& $T_{c0}$&$\TBKT$& $b$ \\ 
	        &   nm& $\Omega$ &nA& $\Omega$&K&mK&  \\ \hline
  J200-1   &  200  & 720 &4& 8.5 & 0.23  &     &  \\ 
	J200-2   &  200  & 425 &5&      &       &     &  \\ 
  J200-3   &  200  & 410 &10  & 1.4&0.35   & 130  &  6.1 \\ 
  J200-4   &  200  & 470 &   50  & 1.0&0.5    & 135  &  8.6 \\ 
  J300-3   &  300  & 370  &   30  & 1.3&0.38   & 175  &  7.2 \\ 
  J400-1   &  400  & 650  &   0   & 16.0&0.285  &     &  \\ 
	J600-1   &  600  & 440   &   0   & &       &     & \\ \hline

\end{tabular}
\end{table}


 \subsection{BKT incipient transition}

In superconducting films, below the mean field temperature $T_{c0}$, the pairing amplitude is finite but overall superconducting phase coherence cannot be established due to thermal fluctuations.  By  further lowering the temperature, we enter a crossover region towards a BKT transition typical of  dirty  thin films. The low temperature behavior of $R(T)$ deviates from the paraconductivity  regime which is power-law like and enters an exponential-law behavior as shown in Fig.~\ref{fig_AL_BKT}b.     
 At $ T_{c0} > T >  T_{BKT} $, the global phase coherence is destroyed by thermally induced phase fluctuations  in the form of free vortices which  produce dissipative conduction due to a finite flux flow resistance.\cite{Berezinski1972JETP,Kosterlitz1973JPC}. In our case,  
 the wide graphene sheet, as well as the large overlap area between the graphene and the Al/Ti pads, provides enough space for hosting even extremely extended vortices (such as Pearl vortices appearing in very thin films). The fit of the measured  $R(T)$  by using  the interpolation formula quoted by Halperin and Nelson\cite{Halperin2009JLTP}, which is valid for $T > \TBKT$:
 \bea
 [ R(T)]^{-1} = \frac{0.37}{b} \left [R_N\right ]^{-1}  \sinh^2 \left [ \left (\frac{b t_c}{t} \right )^{1/2}\right ] ,
  \label{halpNels}
  \enea
is quite successful over more than one decade of resistance values ({\sl  thick broken lines in Fig.}\ref{fig_AL_BKT}b) and covers the full  temperature  crossover  including  the AL  regime. Given $ T_{c0}$ as  extracted from Fig.\ref{fig_AL_BKT}a, the two fitting  parameters here used are  $\TBKT$ and $b$. In Eq.\ref{halpNels}  $t_c = (T_{c0} -\TBKT ) / \TBKT$ and $ t = (T-\TBKT ) / \TBKT  $ appear.   The $0.37$ prefactor is chosen by  Halperin and Nelson to match with the AL linear dependence  of Eq.\ref{ro} with  $R_0 =  16 \hbar /e^2$. Although our fitted values of $R_0$ are at least one order of magnitude smaller than this value derived rom the Ginzburg-Landau theory, we have kept it unaltered as the  fit appears to be rather insensitive to it. 
 The dimensionless parameter $b$  is related to the ratio between the loss in condensation energy at a vortex core  and the superfluid stiffness. It is remarkable that the values of $b$ reported in Table~\ref{table1} are of the same order as the accepted values for the  2D-XY model \cite{Fiory1983PRB} in Indium Oxide films.
%
%
The fitted temperatures $\TBKT$, of the order of  0.1~K, as well as the $b$  and $R_0$ values  are reported in Table \ref{table1}.
%

\section{Hysteretic magnetic field dependence of the Josephson current}
\label{sec_hysteresis}

\begin{figure}
\includegraphics[width=0.95 \linewidth]{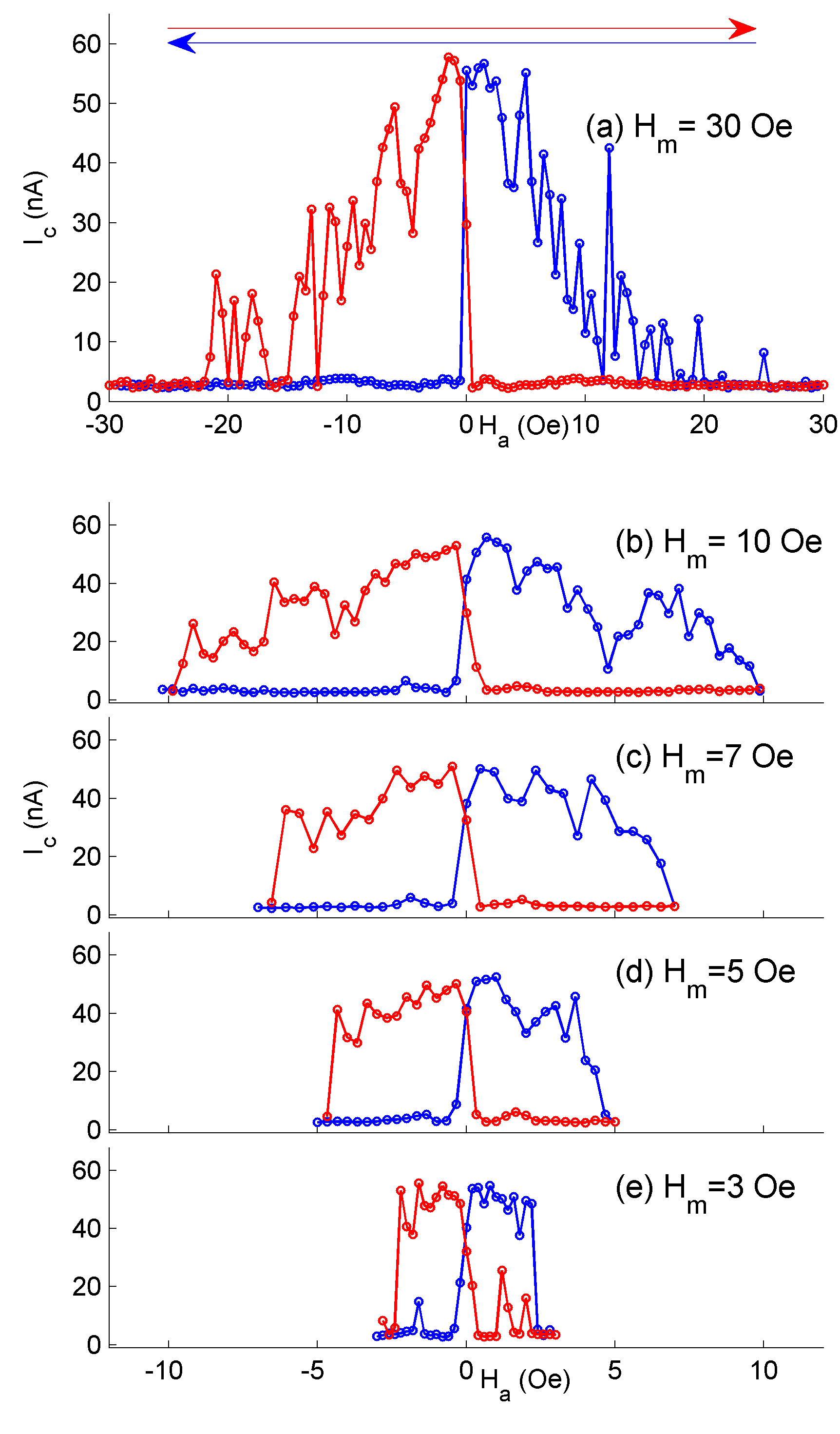}
\caption{{ Critical Josephson current $I_c$ as a function of the applied magnetic field $H_a$ for the J200-4 junction.} In all panels, the red (blue) color refers to sweeps performed from negative to positive (positive to negative) magnetic fields. Magnetic sweeps range up to $H_m$ = 30 Oe ({a}), 10 Oe ({b}), 7 Oe ({c}), 5 Oe ({d}), 3 Oe ({e}). The arrows on top indicate the direction of the magnetic field variation.}
\label{fig_butterflies}
\end{figure}  

Cooling the junctions with $ L <  400 $ nm down  to $T = 280$ mK  in ZFC, a Josephson current is established, with $I_c \approx 50$ nA, notwithstanding the fact that there are thermal fluctuations which produce free vortices and antivortices  in the graphene sheet ($T > \TBKT$).  In Fig.~\ref{fig_butterflies} we report various hysteresis loops of $I_c$ with an externally applied magnetic field  $H_a$  for the  junction J200-4 of smaller area  ${\cal{A}}_{J200} \simeq$ 1~$\mu$m$^2$. For this junction, $L = 200$ nm. 

The sweeping of the applied field $H_a$ is:  
$H_a=-H_m \to H_m$, ({\sl red  curve}) and  
$ H_a = H_m \to -H_m  $,  ({\sl blue  curve}).
Here  $H_m$ is defined as the maximum attained $|H_a|$ field before initiating the decreasing in the sweep.  
We find that  $I_c$  rapidly drops to very low values with increasing  field $|H_a|$ in both positive and negative directions
of the sweep.
%
By contrast, in decreasing $|H_a|$, $I_c$  appears to recover and is strongly sensitive to magnetic field variation displaying fluctuations between various runs (see  {\it e.g.} Fig.~\ref{fig_butterflies}a, close to  $H_a \simeq 10$~Oe). 
This is a robust feature, as demonstrated by the various measurements of the same junction presented  in Fig.~\ref{fig_butterflies}a--e in which  $H_m$ is varied.

The junctions reported in the present work are characterized by non-hysteretic IV curves with a finite slope in the superconducting branch, as shown in Fig. \ref{fig_VI}. In the framework of the RSJ model, this phenomenology could be explained by considering diffusion of the phase particle along the washboard potential, when the Josephson energy is of the same order of the thermal energy \cite{barone}. Indeed, phase diffusion has been observed recently in graphene-based JJs \cite{Finkelstein}. Nevertheless, the measurements in presence of applied magnetic field clearly indicate that phase diffusion is not the main dissipation process. Indeed, the magnetic field changes the amplitude of the supercurrent (estimated by RSJ fit) and modulates the finite slope of the supercurrent branch in a very anomalous way, since it depends on the magnetic field sweep direction. Modulation of the critical current and of the finite slope of the supercurrent branch are shown in Fig. \ref{fig_butterflies} and in panel (a) and in panel (b) of Fig. \ref{ici}, which reports the dependence of the resistance at zero bias as a function of magnetic field. These last experimental observations cannot be explained within the RSJ model, and within any kind of phase diffusion process \cite{kautz,luigi,davide3}. Their interpretation will be addressed in Section \ref{collapse}.

In a magnetic field parallel to the graphene flake we have measured the same hysteresis on a magnetic field scale enlarged by a factor $\simeq 100$. As some undesired tilting of the sample cannot be excluded, we conclude that the hysteresis is generated by a small spurious orthogonal $H$ component due to a misalignment of the coil in the parallel geometry. This confirms that the electronic properties of the system are not appreciably affected by a field $H_a$  parallel to the flake, while the orthogonal component of the field is the main actor.
We can also exclude magnetization effects of non-superconducting origin in the Al contacts. Indeed, in the normal phase ($T>T_c^{Al}$) we do not measure  any hysteresis. The hysteresis in the supercurrent is strongly dependent on the geometry of the weak link, as, by excluding the weak link, and contacting one of the Al islands alone, no hysteresis appears  in the supercurrent.

\section{Collapse of the Josephson supercurrent close to zero magnetic field }
\label{collapse}

\begin{figure}
\begin{center}
\includegraphics[width=0.9 \linewidth]{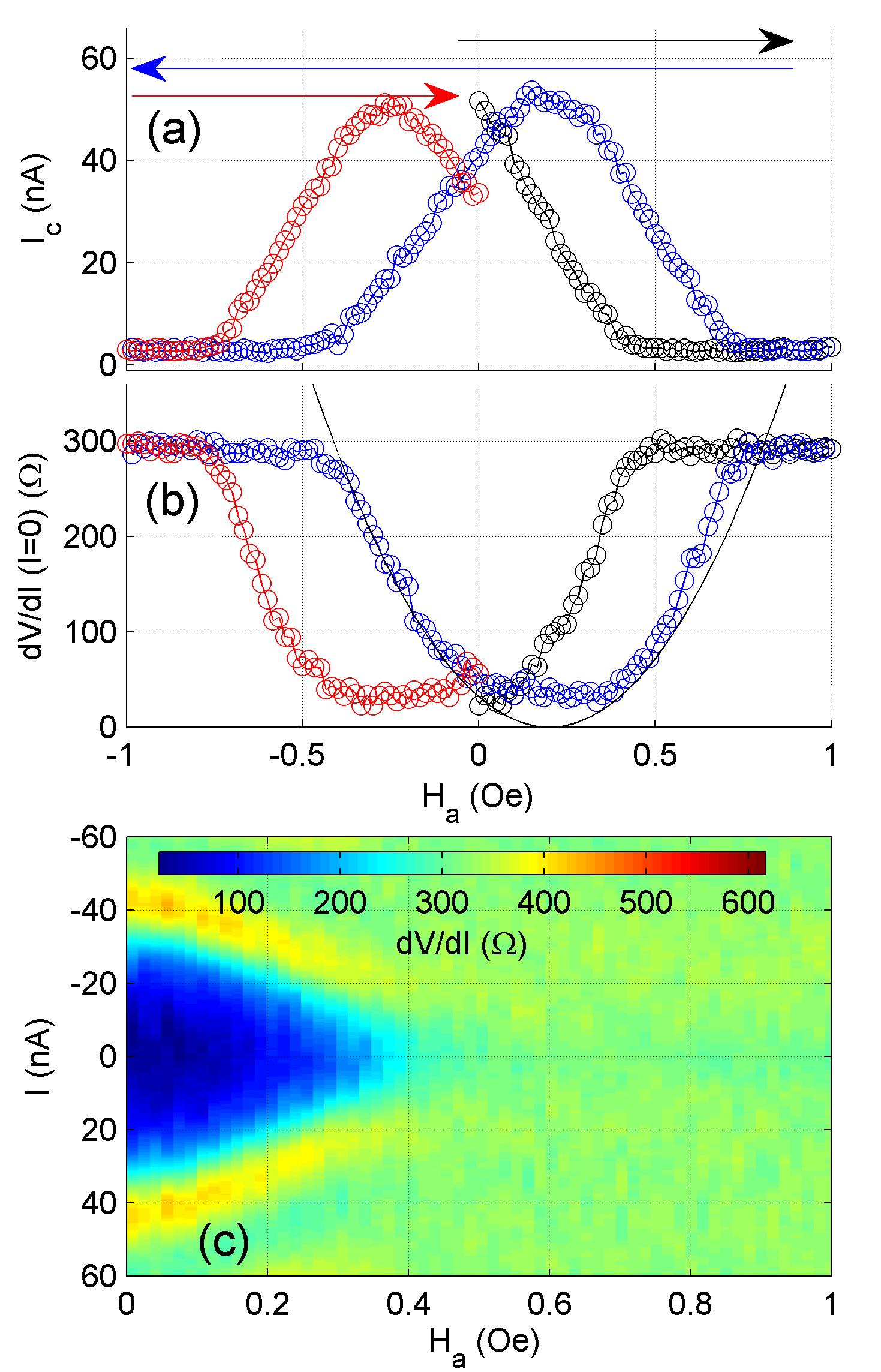}
\caption{{Zoom on the hysteresis of Fig.~\ref{fig_butterflies}  at low magnetic fields ($H_m \le 1$~Oe) for the J200-4 junction, taken at $T=$ 280 mK.} 
{ (a)} Estimated critical current  during a magnetic sweep (sweeping direction given by the arrows). 
 {\it Black circles:} 
the sweep  starts at $H_a=0$ Oe  at ZFC and goes up to $H_a$= 1~Oe. No  flux is trapped initially in this first sweep after the  cool down. 
{\it Blue circles:} the magnetic field is swept from 1~Oe  to -1~Oe. 
{\it Red circles:}  the magnetic field is swept back from $-1$~Oe  to 0~Oe.
{ (b)} The differential resistance at $I=0$
recorded when the magnetic field is swept from 0 to 1~Oe ({\it open dark circles}),
from 1 Oe  to -1 Oe ({\it open blue circles}) and
from -1 Oe to 0 Oe ({\it open red circles}).
The black curve is a parabola, which serves as a guide for the eye.
{ (c)} Colormap of the differential resistance $dV/dI(H,I)$ recorded during the  first sweep after ZFC, from 0 Oe  to 1 Oe. It corresponds to the {\it black circles} of panel { a}, {\it i.e.} it reports the first collapse of the critical current $I_c$. 
For $H_a > $ 0.4 Oe , viscosity $\eta \propto H $  [see Eq.~\ref{vivi}] and  the Critical State Model applies, with $R \simeq 300$~$\Omega $.}
\label{ici}
\end{center}
\end{figure}

In this Section,  we propose an interpretation of  the unusual collapse of the Josephson critical current $I_c$, close to zero field, as well as of  the revival at the inversion point.
$I_c(H_a)$ and the corresponding $R(H_a)$ are reported in Fig.~\ref{ici}a,b
at a magnetic field scale much smaller than the one appearing  in Fig.~\ref{fig_butterflies}.
In Fig.~\ref{ici}a, the first drop of $I_c$ with the increasing of $H_a$ after ZFC is marked by open black dots. From this curve we extract the magnetic field value $H_f$ = 0.45 Oe at which the $I_c$ collapse is  completed.
Furthermore, 
the maxima of $I_c$ (and the minima of $R$) in the back sweeps are
shifted to $|H_a| \simeq 0.25$ Oe, where the sign depends on the sweep direction.
Additionally, comparison of  Fig.~\ref{ici}a with  Fig.~\ref{ici}b   reveals that in all the  magnetic field sweeps there is a direct relation between $I_c$ and the residual magnetoresistance $R(H)$.

The hysteresis observed in Fig.~\ref{ici}a,b shares two important features with the hysteresis which is commonly observed in the magnetoresistance of granular superconductors:

{\sl i)} for the same value of $H_a$, the resistance in the decreasing $|H_a|$ curve
is lower than that in the increasing $|H_a|$ curve; 

{\sl ii)} the minimum of $R$ is obtained before $|H_a|$ reaches zero in
decreasing $|H_a|$.
The usual model to explain these features in granular superconductors
is the so-called two-level critical state model, developed in Ref.~\onlinecite{Ji1993PRB},
in which superconducting grains 
trap and pin vortices, inducing an hysteresis, whereas the vortex dynamic at the grain
boundaries gives rise to a finite resistance.
In our view, pinning centers are present under the Al pads
and
the observed incipient BKT transition signals the presence of vortices in the same area.
%
In the simplest interpretation, the areas under the Al pads correspond to two separated grains, whereas the bare graphene junction  corresponds to the boundary between these two grains.
We cannot exclude a more complex picture, in which the areas under the Al pads are themselves constituted of several smaller grains.

Coming back to Fig.~\ref{ici},
we try now to understand why the collapse of $I_c$ with increasing $|H_a|$ is so drastic.

A viscosity of the vortex liquid can be extracted from the magnetoresistance. 
Let us assume that the bias current density $\vec{ j}_{ext} $ flows in the $\hat x $ direction across the weak link of width $W$. Phenomenologically, Lorentz force drags  flux lines  moving with velocity $\vec{v}_L$ and  viscosity $ \eta $ along the $\hat y $ direction.  For a viscous inertial flow in a homogeneous film of thickness $d$, the magnitudes of these vectors are related by: 
   \beq
     j_{ext} \:  \frac{ \Phi_0}{ c }   = \eta \: v_L .
     \label{drift}
 \eneq  
$\Phi_0= hc/2e$ is the flux quantum. The flux flow resistivity  $ \rho_f  = R W /L $ is related to the viscosity by\cite{Ji1993PRB,tinkham}:
  \beq
\rho _f =  \frac{E}{ j_{ext} }  = H \:\frac{ \Phi_0}{ \eta c^2 },
\label{vivi}
 \eneq
 where the compensation of the drift and the Lorentz  force, $ \vec{E} = \vec{H} \times \vec{v} _L/c$, for an inertial  vortex flow,  have been used in the second equality.  

In conventional type-II superconducting films, the  magnetoresistance $R(H)$ is linear with $H$, what tells that $ \eta $ is constant with $ H$. 
This is not the case here for $|H_a| < H_f$, as shown in  Fig.~\ref{ici}b. In this range of fields, 
 a parabola  provides a rather good  fit of  $R(H)$.
Beyond  $H_f$, there is a  change of behavior and $R(H)$ becomes constant with the field.
 This implies that there is a regime of high viscosity and small resistance, $ \eta \propto 1/H$, for $|H_a| < H_f$ where $R(H)$ is roughly parabolic, and a regime of constant resistance with  $\eta \propto H$  for larger applied fields.  As, by increasing $H$, we expect that the density of unbound vortices increases, the crossover in Fig.~\ref{ici}b can be most likely attributed to a change in the dominant  interaction between vortices. 

To support this assumption, we have performed a classical simulation of 2D disks interacting via a long range Gaussian repulsive force and a short range quasi-hard core force.
The details on the classical simulation can be found in Appendix A. Such 2D classical simulations have been used in the context of the BKT phase transition\cite{Prestipino2011PRL} and the melting of a quasi-3D-vortex-glass  with increasing $H_a$ at fixed temperature has been observed in YBCO\cite{Koch1989PRL,tinkham}, though at higher magnetic fields. This is not a proper melting, because  a vortex-glass phase  is not expected to take place  at finite temperatures in a 2D layered structure  as is the one discussed here.

Our simulation shows that a rather rigid gossamer-like texture forms at low $H$, when the density of vortices is rather low and the long range repulsion is dominant. 
We neglect disorder in our simulation, because an extended rigid texture with long range correlations cannot be pinned by the random configuration of pinning centres  expected to be present under the Al pads. This would not be the case at higher temperatures because thermal fluctuations would soften the texture. 
It follows that the vortex structure can drift freely when the applied current acts as a force on it providing the dissipation mechanism. Numerically, we find  $\rho_f \propto H^2$ at low $H$.
%
By contrast, when $H$ increases, the density of free vortices increases, the long range correlation looses its dominant role  and the texture starts melting. In this regime, our simulation shows a saturation of the resistivity.
Therefore, our simulation reproduces  the observed magnetoresistance thus confirming our interpretation.
Needless to say, all these features are a unique property of the geometry of the device and a straightforward consequence of the incipient BKT transition in the graphene sheet and cannot be found in thicker weak link films. 

 \begin{figure}
\begin{center}
\includegraphics[width=1 \linewidth]{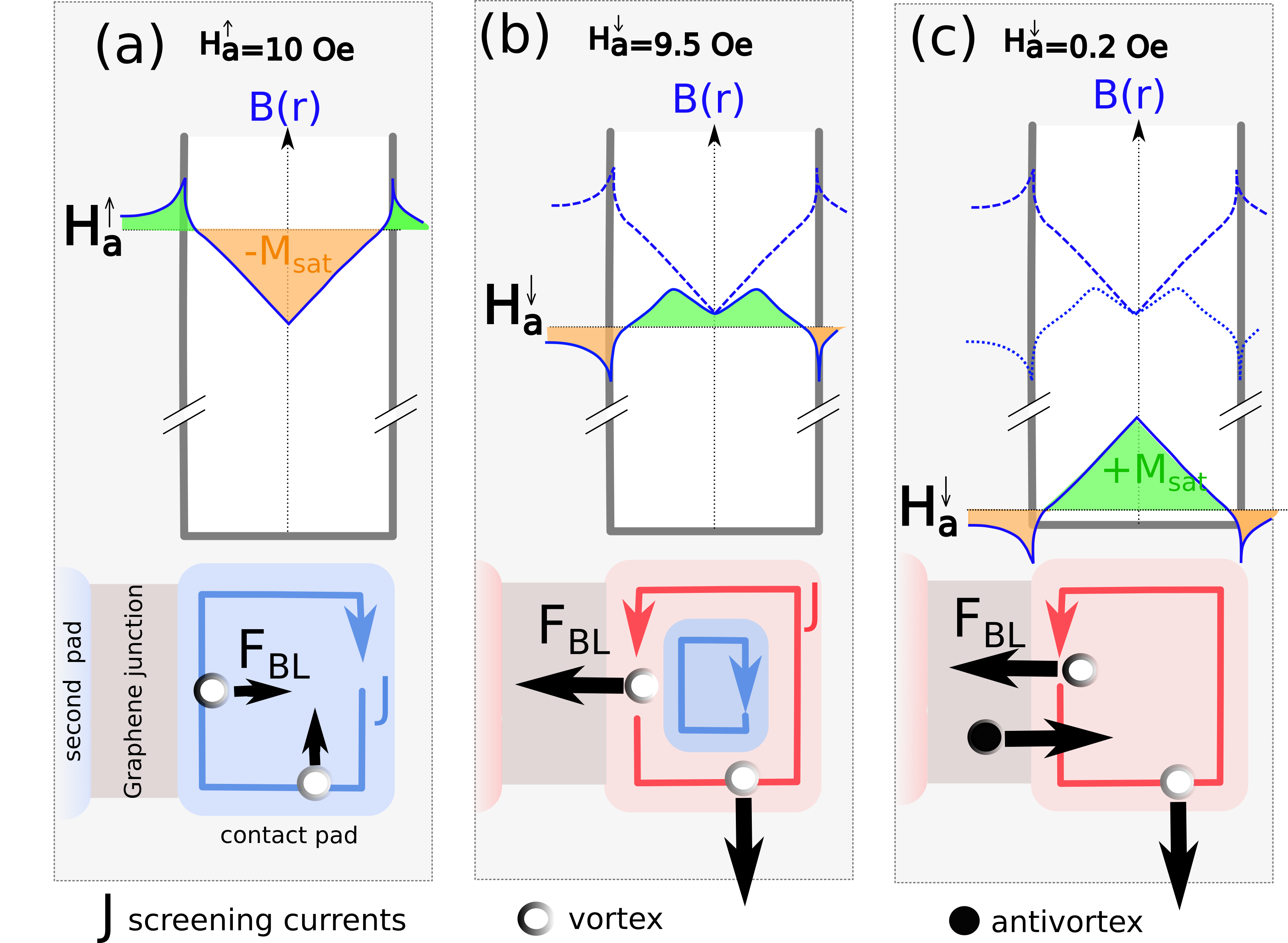}
\caption{{Sketch of the vortex dynamics in the hysteresis of the Josephson current $I_c$ according to the CSM.}
The contacts are considered as grains. Just one of the contacts of the junction is drawn. {\it Top:} applied magnetic field $H_a$ and field induction $B(r)$  in the interior of the contact ($B=0$ is at the bottom of the well). {\it Bottom:} Flowing screening currents densities $J_c$ ({\it dark blue } and  {\it dark red}) and vortex  ({\it white circle}) or  antivortex ({\it black circle}) motion under the action of the Bean Livingston  force, $F_{BL}$, which is due to the energy gradient named "Bean--Livingston Barrier''. 
{ (a)}  $H_a \simeq H_m$: vortices enter the Al pads in increasing $|H_a|$.
{ (b)}  $H_a < H_m$: in decreasing $H_a$, the current density $J_c$ is reversed at the boundary and  
the Bean Livingston Barrier is washed out, so that vortices exit from the pads. Since $H_a^{\downarrow} $= 9.5 Oe $>H_m-2H^*$, an inverted cusp at the
center of the magnetic field profile can be observed ($H^*$  defines the field at which the contacts are fully penetrated, {\it i.e.} the central cusp  still touches $B=0$ in { a})).
{ (c)} $ H_a \simeq  0 $: vortices
continue leaving the contact, while a negative return magnetic field builds up at its edges, which implies that  some antivortices enter the contact  and may annihilate with some of the exiting vortices.}
\label{csmod}
\end{center}
\end{figure}
 
Having interpreted the quick collapse of $I_c$ when $|H_a| $ is turned on, we now focus on  the revival of the Josephson current which is observed in Fig.~\ref{fig_butterflies} when $|H_a|$ is decreased.
As mentioned already, our system can be regarded as granular,  like a coated-conductor\cite{Palau2004APL,Palau2007PRB}.
In the simplest view, 
the grains correspond to the graphene area overlapping with the Al pads, and the grain boundary,
across which  the dissipative/non-dissipative Josephson supercurrent flows, 
corresponds to the bare graphene weak link.  
In the  Critical State~\cite{Bean1962PRL,Zeldov1994PRB} which forms in the "grains", 
vortices nucleate at the edges of the grains  
when $|H_a|$ increases
and they attempt to move inside, toward the grain interior.  Since pinning forces are opposed to vortex diffusion, a magnetic field gradient is formed in the grain
with its resulting current profile. 
The recovering of the Josephson current when the sweeping of the magnetic field is inverted, can be interpreted by considering  the vortex dynamics in the
grain region, see Fig. \ref{csmod}. Let us consider the $H_a > 0 $ case for sake of the discussion. The same can be argued for the $ H_a< 0 $ sweep.
In increasing $H_a$, vortices continuously enter the grains overcoming the Bean-Livingston barrier\cite{Bean1964PRL,deGennes}. This barrier is the sum of the contribution of the screening current (whose sign
can be positive or negative depending if $|H_a|$ increases or decreases)
and the image force (which does not depend on the direction of the magnetic field sweep). Vortices entering the grains  produce  penetration of the magnetic flux inside them. Due to the unequal diffusion inside the pad area, the field acquires a slope inside, which is approximately uniform according to  the Bean CSM (see Fig.~\ref{csmod}a). 
At the very first moment when  the magnetic field starts being reduced, those vortices which are loosely pinned to the defects are 
expelled immediately from the grains and swept away along the graphene "grain boundary'', crossing the path of the Josephson current (Fig. \ref{csmod}b). 
This is just the starting
moment for the recovery of $I_c$. Further reduction of $H_a$ generates an inversion of the magnetic field gradient which penetrates the grains. This  is accompanied by the inversion of the flow direction of  the critical screening  current at the  grain boundaries, which, in turn,  lowers the Bean-Livingston barrier\cite{Bean1964PRL,deGennes} for   vortex flow out of the grains. 

The Bean-Livingston barrier is a known source of anomalies in the magnetization curves of type-II superconductors\cite{Kopylov:1990}, because the barrier profile in the vicinity of the grain boundary differs in increasing or decreasing $|H_a|$.

Beyond this point, several tentative explanations can be proposed.

{\sl i)} only vortices which are depinned move along the grain boundary and are expelled. Vortices which remain pinned inside the grains do not contribute to the resistance.  Fig. \ref{csmod}c sketches what happens when sweeping $H_a$ down to zero field. 

{\sl ii)}  crowding of the vortices  ejected out of the grains in the weak link channel with core repulsion between them can strongly increase viscosity in the channel and reduce their flowing across, so that the related  flux flow resistance is also drastically reduced.

{\sl iii)}   annihilation of some of the exiting vortices  by antivortices of the broken $v$-$\bar v$ pairs created by thermal fluctuations or generated by the inversion of the local magnetic field at the grain boundary. All of this gives rise to the  recovery of  $I_c$  in decreasing $H_a$.

Let us comment now in more details why
we observe that the maximum of $I_c$ is shifted to positive $H_a \approx 0.25$ Oe, when sweeping from positive fields 
down to zero field. 
According to the CSM, decreasing $H_a$ leaves a negative residual  magnetic field $H_r^{Al}$
at the edges of the grains, 
which arises from the  trapped magnetic field that survives and from the demagnetizing factors in the grain (see the magnetic field profile in Fig.~\ref{csmod}c). 
The local magnetic field in the weak link $H_{loc}^{graphene} =0$, 
at which the system presents the maximum of $I_c$,
corresponds to  the field at which $H_a = -H_r^{Al}  $, because
$ H_{loc}^{graphene} = H_a+ H_r^{Al} $. We performed magnetic fields sweeps with various $H_m$ in the range 0.5--3~Oe
and found that $H_r^{Al}$ saturates at 0.25 Oe when 
$H_m \geq 1$~Oe.

If  the cycle is continued and  the magnetic field  is reduced beyond   $ H_{loc}^{graphene}=0$, we find  the  collapse of $I_c$ once  more and we enter the region of  negative  $H_a$ values ({\it blue curve} with increasing  $|H_a|$ in Fig.~\ref{fig_butterflies}). 
 
Within the  CSM, we can relate the value of $H_r^{Al} = -0.25$~Oe to the field at which the Al pads are fully penetrated by the magnetic field, conventionally  denoted by  $H^*$.  This is  the magnetic field at which we can assume that most of the pinning centers in the Al pads  have captured  a flux line. 
Following  Refs. \onlinecite{Palau2004APL,Palau2007PRB}, we estimate
\beq
  H^* =  H_r^{Al}\: \frac{n\: a }{xt},
\eneq
where $x$ and $n$ are numerical  dimensionless demagnetization factors and $t$ and $a$ correspond to the thickness and to the linear size of each of the Al pads, respectively. In a first approximation, we have considered $a =4\, \mu m$ and we take $t$ as the thickness of the Al pad ($t=80$~nm) .
This provides a value of $H^*=0.5$~Oe just beyond the field
$H_f \simeq  0.45$~Oe
at which the collapse of $I_c$  is completed and the Critical State is fully established. Therefore, for fields higher than $H^*$, a tiny Josephson current coexists with a sizeable dissipation induced by  the flow of free vortices. 

Experimentally, the hysteresis can be observed up to magnetic fields as high as 20 Oe (see Fig.~\ref{fig_butterflies}a). 
At higher magnetic field, the $I_c(H)$ is reversible. This suggests that  20 Oe corresponds to the irreversibility field, {\it i.e.} the field at which the hysteresis and the  Critical State picture  disappears, since the vortex lattice fully liquefies.

 \section{Fraunhofer pattern of the Josephson Junction}

 \begin{figure}
\includegraphics[width=0.9 \linewidth]{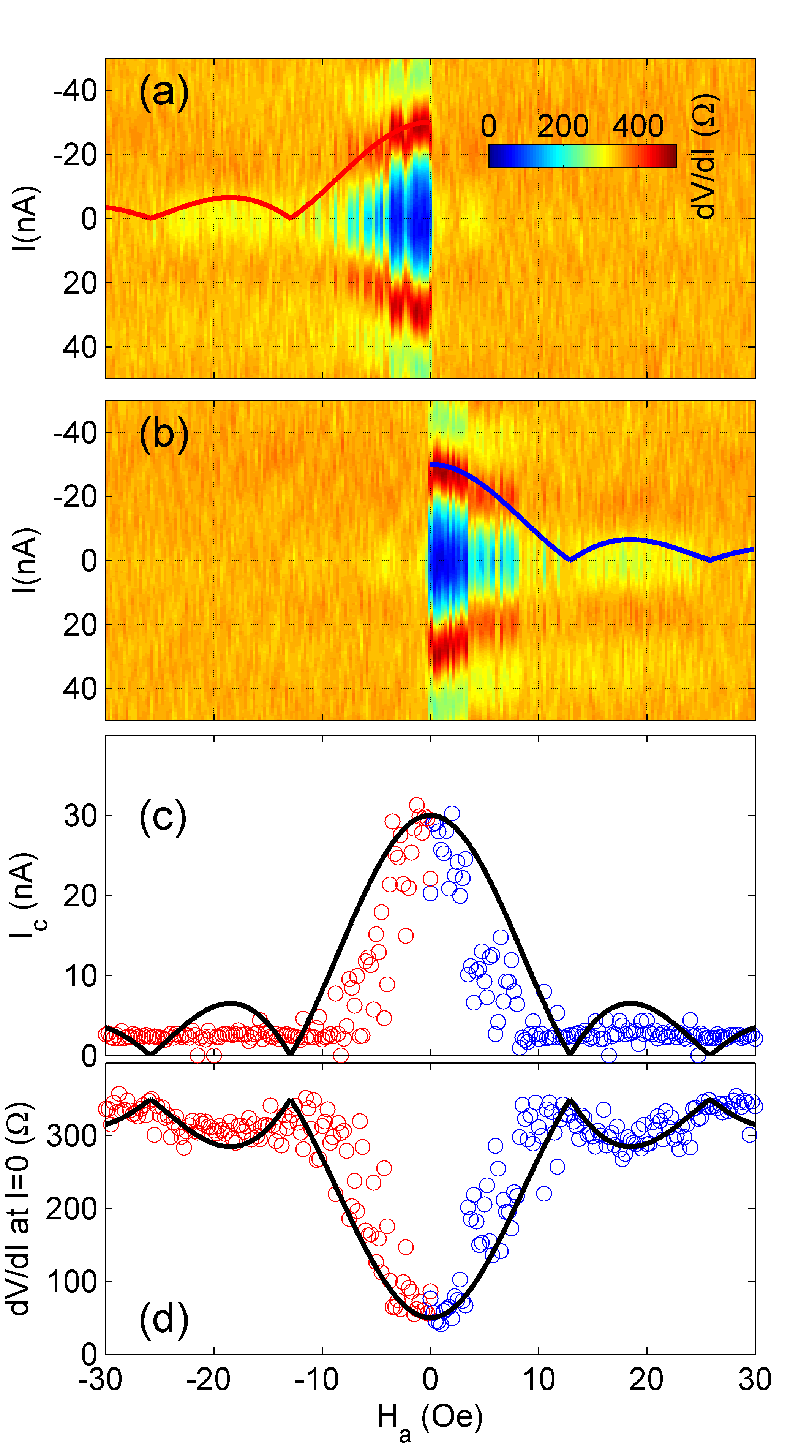}
\caption{{The Fraunhofer pattern in junction J300-3.~}{ (a), (b)} Colormaps of the differential resistance $dV/dI(H,I)$ for the junction J300-3, evidencing a large hysteresis.
 { (a)} The magnetic field is swept from -30 to 30 Oe. \\
 { (b)} The field is swept from 30 to -30 Oe.
The blue area corresponds to the low resistive region. 
The superposed blue and red curves are a Fraunhofer interference pattern, given as a reference, corresponding to a total area $S_\mathrm{eff}= 1.6~\mu$m$^2$.\\
{ (c)} Critical current as a function of the magnetic field ({\it red open circles:} sweep from -30 to 0 Oe ; {\it blue open circles:} sweep from 30 to 0 Oe). The black curve is the theoretical Fraunhofer pattern fit calculated using the RSJ model as explained in the main text.
{ (d)} The residual resistance at $I$= 0 nA ({\it red open circles:} sweep from -30 to 0 Oe ; {\it blue open circles:} sweep from 30 to 0 Oe) also reproduces the same Fraunhofer pattern, indicated by a thick black line as a guide for the eye.}
\label{fig_colormapJ300}
\end{figure}  

In Fig.~\ref{fig_colormapJ300}a,b we map the differential resistance $dV/dI$  of the J300-3 sample as a function of $H_a$ and $I$, for the two directions of the magnetic field sweep.
The dark blue areas correspond to differential resistances $dV/dI$ below 5 $\Omega$. 
The full curves in  panels a--d are guides to the eye corresponding to  the 
 Fraunhofer pattern usually appearing in extended junctions with uniform distribution of the supercurrent density $J_c $, giving $I_c \propto |\sin(\pi B S_\mathrm{eff}/\Phi_0)/ (\pi B S_\mathrm{eff}/\Phi_0)|$
 for an effective area
$S_\mathrm{eff} \simeq  1.6\:\mu$m$^2$. 
In panels a and b, this Fraunhofer pattern fits roughly the experimental data obtained in decreasing $|H_a|$. 

$I_c$, as obtained through the RSJ fit of the I-V curves (see Section \ref{sec_samples}),
 is reported in Fig.~\ref{fig_colormapJ300}c
as a function of the applied magnetic field.
By this method, the Fraunhofer pattern is less visible, while it is better  
 retrieved in Fig.~\ref{fig_colormapJ300}d,
which shows $dV/dI(H_a)$ at $I=0$ from the I-V characteristics.
The little shift of  $ H_{loc}^{graphene}=0$ discussed in the previous Section cannot be appreciated on the magnetic field scale adopted here. 
%

\begin{figure}[h]
\begin{center}
\includegraphics[width=0.95 \linewidth]{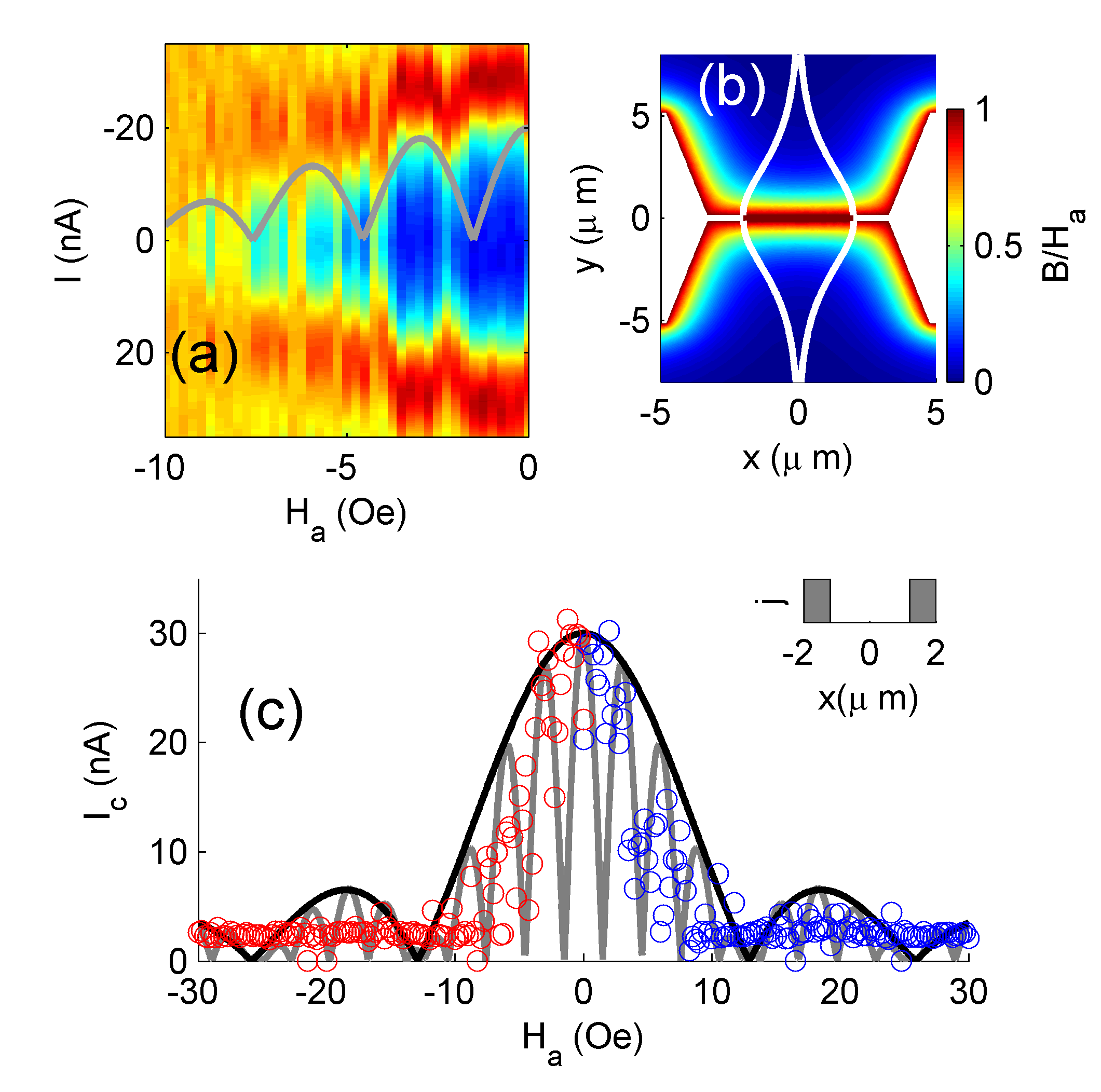}
\caption{{The Rosenthal model.~}{(a)} Zoom of the colormap of the differential resistance $dV/dI(H,I)$ for the junction of width $L\approx 300$ nm presented in Fig.~\ref{fig_colormapJ300}, with the same color code. 
The data have been collected by decreasing the absolute value of the applied magnetic field. 
The dark blue area corresponds to the superconductive region. 
The superposed gray curve is a Fraunhofer interference pattern  given as a reference, corresponding to an effective area $S_\mathrm{eff}^R\simeq 9 \: \mu$m$^2$.
 {(b)} Colormap of the field penetration within the contacts, for a Pearl penetration length $\lambda= 1 \:\mu$m. The white line is a path enclosing the graphene junction which is perpendicular to the current flow.
{(c)} Magnetic pattern. The fast oscillating $I_c$  {\sl in grey} curve,  as derived from the model in  (b), with an inhomogeneous current density distribution $J_c$ (sketched in the inset),  is added to the  Fraunhofer pattern fit reported  in Fig. \ref{fig_colormapJ300}c. The current density distribution  $J_c$ adopted in the fit is concentrated at the edges of the junction within two 200 nm wide  strips. The same oscillations appear in (a) on a reduced scale.}
\label{fig:FP}
\end{center}
\end{figure}

 
In Fig.~\ref{fig:FP} we zoom in the down sweep map of Fig.~\ref{fig_colormapJ300}a and concentrate on the differential resistance. It appears clearly that the Fraunhofer oscillations considered up to now are just the envelope of a much faster  oscillation pattern.  This pattern is not strictly periodic  with a pseudo-period of 1-3 Oe (see Fig.~\ref{fig:FP}a). Occasionally, disturbances  as jumps of the measured residual resistance at  $ I=0 $, can be also spotted, probably due to flux jumps. This confirms that some vortex dynamics is taking place. We are unable to keep track of these  microscopic irregular features, but we have set up a macroscopic point of view to account for the non negligible magnetic field penetration in the Al/Ti contacts. Indeed, the pseudo period of  1--3 Oe  corresponds to a much larger  effective area 
$S_\mathrm{eff}^R\simeq 9 \: \mu$m$^2$ than the weak link itself.  We have calculated numerically the magnetic field profile  for contacts in the thin film limit, following Rosenthal {\it et al.}\cite{rosenthal},  by solving the London equation,  
$\nabla ^2 J_s  -\lambda ^{-2} J_s =0$ in a quasi 2D contact.  
A color map of the penetration of the field in the contacts appears in  Fig.~\ref{fig:FP}b with  a penetration length  $\lambda  = 1 \:  \mu$m and a width of the weak link $W= 4 \: \mu$m. This value of the penetration length is significantly larger than the value usually reported for bulk Al (around a few tens of nanometers) and can be interpreted as a Pearl penetration length.  The phase difference between two points is then calculated in the London gauge, by integrating the vector potential (which is proportional to the current) over a path which links these two points. We choose a path along which the longitudinal component of  $J_s$  vanishes. This is plotted  as the white curve in  Fig.~\ref{fig:FP}b.  The resulting enclosed area  is  
$S_\mathrm{eff}^R\sim W^2 /2 \simeq 8 \: \mu$m$^2$,
which is  very close to the value extracted from the pseudo period. 
 The fast oscillating grey curve  in  Fig.~\ref{fig:FP}c  is obtained by assuming  an inhomogeneity in the critical current density $J_c$. Some inhomogeneity  is expected  {\it e.g.} in planar devices where current focussing   is typically observed in overdamped junctions at the edges of the junction\cite{barone,hart2014Nature}. The current density distribution  adopted  in the fit is drawn in the inset  of Fig.~\ref{fig:FP}c. Both the fast $I_c$ oscillations ({\it full grey curve in Fig.~\ref{fig:FP}c}) and the envelope modulation are retrieved by choosing  $J_c$ concentrated in a strip 200~nm wide, at each of the boundaries of the graphene sheet. 

This  model, which combines the  London equation with an inhomogeneous current flow in a macroscopic approach,  though remarkably sound, cannot reproduce some of  the most puzzling characteristics of the conduction: the aperiodicity of the fast oscillations, the presence of flux jumps, not to speak about the hysteresis.

 \section{Discussion} 
 We have  deposited a graphene monolayer on SiC by CVD and patterned various in-plane Josephson weak links on the same sample, with  Al/Ti thin contacts at a variable distance $ L \gtrsim 200$~nm, as shown in Fig.~\ref{fig:sketch}. Details of the fabrication can be found in Ref. \onlinecite{noi16}.  Each of the junctions  can be thought of as an extended  weak link, $W= 4 \mu$m wide, between two thin  metal grains. The junctions with
$L \simeq $  200--300~nm 
 show Josephson conduction. Our analysis of $R(T)$ allows us to define a mean field critical temperature $T_{c0} $  for our  planar devices in the Aslamazov-Larkin paraconductivity regime  precursive of superconductivity. The  Halperin and Nelson interpolation formula for an incipient BKT  transition  captures the full  temperature dependence of  $R$  down to the operation temperature (280 mK). The value of the dimensionless $b$ parameter which we get from the fit is quite large, of the order of the one found in the original formulation of the transition in the 2D-XY model. Our $b$ value is also close to the one given in Ref. [\onlinecite{Fiory1983PRB}] for Indium/Indium Oxide  granular films. The ratio between the vortex core energy and the superfluid stiffness $ J_s = (\phi_0^2/4\pi )( d/ \lambda ^2 ) $ that we find is $\mu  / J_s = \pi ^2 \sqrt{ b} /4  \sim 5 $ ($d$ is the layer thickness and $\lambda $ is the London penetration length). Such a large ratio is related to the  difference  between $T_{c0}$ and $T_{BKT}$. Their separation is close to one order of magnitude, what is seldom found in NbN or Al films\cite{yongPRB87.184505}. The superfluid stiffness is quite small because the inverse of the Pearl length appears for the graphene sheet, $\lambda_P = \lambda^2 /d \sim$ 300 $\mu$m.
 
  A Berezinskii--Kosterlitz--Thouless transition at  lower temperature,  $T \sim 100 $ mK,  allows to interpret  the apparently odd dependence  of the Josephson current on magnetic field. This can be attributed to the  dynamics of the free interacting vortices hosted by the  graphene sheet which is quite extended under the Al/Ti pads. Vortices  originated by broken $v$-$\bar v$ vortex pairs  penetrate  the overlapping Aluminum pads and can be  pinned by the impurities, including unavoidable defects due to photoresist residue. 
 
  Figs.~\ref{fig_butterflies},\ref{fig_colormapJ300} entail the unique features of these  structures, in which the Josephson critical current $I_c$ is hysteretic,  when cycling with an applied magnetic field $H_a$, after a ZFC. 
  The  Josephson current, though sensitive to the phase difference modulated by the magnetic field,  is always accompanied  by some dissipation. The latter is strong when increasing $|H_a|$, but it is  quite low  when  reverting the  sweeping of the field. We attribute the hysteresis to the presence of  a large number of impurity centers at the Al pads which  pin the vortices that are pushed into the dirty metal. The S-N-S structure resembles a granular material  with the  graphene gap playing the role of a grain boundary\cite{Ji1993PRB}.  The  Al/Ti  pads enter a Critical State  that can be described by the Bean CSM. Decreasing $|H_a|$, 
  loosely pinned  vortices are expelled out of  the Al grains. In addition, reversal of the current at the boundary of the Al pads produces a lowering of the Bean-Livingston barrier for vortex expulsion, thus enhancing the process. This vortex dynamics  implies a reduction in the flux  flow  across the graphene weak link and a recovery of the Josephson critical current $I_c$. 
%
The hysteresis is fully reproducible.

The Fraunhofer-like  pattern for the Josephson critical current can be fitted within the Rosenthal  model. The physical phenomena at the origin of the supposed  inhomogeneous current flow invoked to justify the fast oscillating pattern  with magnetic field may have many different origins. For instance, current focussed at the edges of the junction is typically observed in overdamped Josephson dynamics \cite{barone}. We conclude that, in the granular picture of Section V there are two different length scales: 
a microscopic scale with vortex pinning  and a macroscopic scale defined by $\lambda \simeq  1\:\mu$m, which defines the macroscopic magnetization of the Al pads and the hysteretic behavior of the supercurrent. The fit of the Fraunhofer-like  pattern that we obtain with the  Rosenthal  model is rather satisfactory. 
Still,  field penetration in the CSM and London equation in the Rosenthal model for the weak link are not enough to explain all the features of the experiment.  Additional features like the non periodicity  of the fast oscillations in the Fraunhofer pattern or some occasional flux jumps appearing in the data  cannot be captured by the continuum macroscopic picture.

 A remarkable feature of the magnetic sweep,  when starting from zero field in the 
$H_a \lesssim 0.45$~Oe  narrow range, is the drastic collapse of $I_c$.
 This is a unique property of the geometry of the device (we have checked that point contacts in which $W$ is much shorter, do not show the collapse) and we attribute it to the incipient BKT transition in the graphene sheet. We figure out  that in this range of $H$ values,  the  free vortices  are quite dilute at 280 mK but, as they originate from breaking of $v$-$\bar v$ vortex pairs due to thermal fluctuations, they are  long range correlated.  A gossamer-like texture is created  in the whole extension of the graphene sheet. Such a solid array is rigid because elastic vibrations of the flux lines are expected to be frozen at 280 mK. At the beginning they cannot move freely, as long as  the applied magnetic field  is so weak that there is no unbalance between the number of vortices and  antivortices. However, 
with increasing  $H_a$,  the  Lorentz force induced by the flowing current starts drifting  rigidly the texture in which vortex of one charge prevail, in the orthogonal direction, generating a flux flow resistance. As long as the texture is rigid, it cannot get pinned by the impurities, because of incommensurability  between the random space distribution of pinning centers and the vortex texture. Indeed, the pinning force vanishes in the average. The flux flow produces the collapse of $I_c$. 

By increasing the density of free vortices, the effect of the long range correlations is reduced and the liquefaction of the texture starts.  In Fig.~\ref{ici}, we are monitoring a  ``weak first order transition'' induced by the quasi 2D contacts\cite{Koch1989PRL}.

 We have simulated this dynamics with the classical diffusion of disks interacting via  a long range potential. It is important to stress that this transition occurs with increasing density of free vortices and has no connection with the melting transition that occurs with temperature in HTc anisotropic superconductors. The latter  is due to thermal vibration of flux lines\cite{tinkham}. Vortices in the  liquid can  feel the pinning forces individually  and give origin to the Critical State and to the hysteresis. 
The energetics of vortex lines  entering or exiting a bulk  Type II  superconductor has been extensively considered in the past, in connection with the determination of $H_{c1}$\cite{deGennes,Bean1964PRL,Aladyshkin2009} and goes under the name of "Bean-Livingston barrier'' for vortex penetration in the bulk of a  superconductor. We have calculated the Bean-Livingston barrier by solving the London equation in the planar structure in the framework of the Rosenthal model~\cite{rosenthal} by adding the vortex and its image and we have found that the barrier is rather small  in our planar structure because it is thin and screening is quite low.  Moreover, the barrier disappears when the current is reversed at the boundary and this has the consequence that the recovery  of $I_c$, when the field sweeping is inverted, is quite fast.  

\begin{figure}
\begin{center}
\includegraphics[width=8cm]{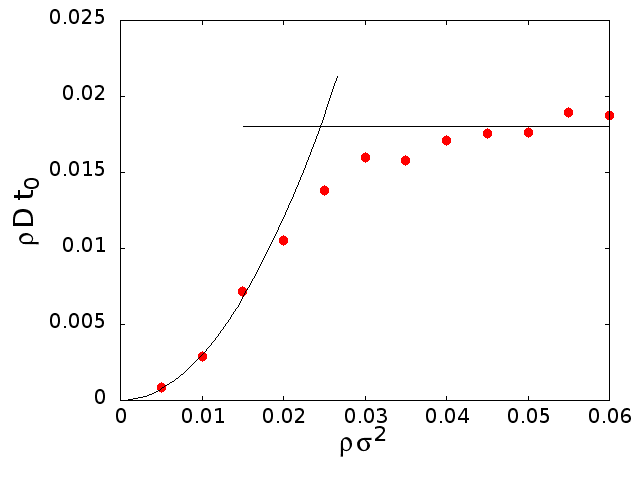}
\end{center}
\caption{Plot of $\rho Dt_0$  as a function of $\rho\sigma^{2}$ for a classical Brownian motion of interacting disks with interaction potential given  by Eq. (\ref{poi}). By trading the model for an analogy with the dynamics of the vortex liquid  at low temperature, this is a plot of the resistivity  versus magnetic field (i.e. $ e^2\rho_f  /\hbar d $ vs. $H \xi ^2/\phi_0 $, where  $d$ is the thickness of the weak link layer). 
}
\label{fig:disks}
\end{figure}

 \section{Summary} 
To sum up, we have measured diffusive transport in a  coplanar graphene Josephson Junction with a  single layer (or, locally, very few) graphene sheet extending under the Al contacts on an insulating SiC  background. The Josephson current, the differential resistance and the response to the magnetic field has been measured down to 300 mK. We find:

a) $R(T)$ curves which can be fitted with the celebrated  formula derived by Halperin and Nelson for a  temperature range in the crossover from paraconductivity  below 1 K, down to Berezinskii-Kosterlitz-Thouless superconductivity with a critical temperature T$_{BKT}$ of the order of 0.1~K (see Fig. \ref{fig_AL_BKT}).

b) The parameters deduced from the fit in the BKT picture suggest that the condensation energy lost in the vortex core is more than five times larger than the phase superconducting stiffness (which is estimated from the screening length $\lambda >$ 1 $\mu$m).

c) Phase coherent Josephson conduction is measured, provided the separation between the Al banks is $<$ 400 nm, although in presence of a small resistance.

d) An unexpected fully reproducible magnetic hysteresis is found: the Josephson current  $I_c(H)$ collapses with $H$ up-ramping as soon as the field is turned on, and undergoes revival as soon as the ramping of $H$ is inverted (see Fig. \ref{fig_butterflies}). For higher fields ($H$ $>$ 20 Oe), the sweeping is reversible.

e) The plots $I_c$(H) are symmetrical and $I_c$ oscillates in the revival, in a way that recalls the Fraunhofer interference pattern.

f) An interpretation of the hysteresis and of the flux flow resistance is given within the Critical State Model, which rests on vortex dynamics with their penetration into the Al banks.

Finally, this work demonstrates that  CVD on SiC  can provide decisive progress towards scalability of superconducting graphene junctions with immediate applicative impact. Unusual properties of the planar measured structures originate  from the superconductive proximity involving  graphene. This  study paves the way to the design of devices having  exactly  the same barrier quality together with constructive parameters that can be selectively changed, in search for the desired  functionalities. This is the only possible path for a hybrid graphene/superconductor technology.

 {\it Acknowledgments:} Enlightening discussions with L. Benfatto, V. Bouchiat, P. Brouwer, J.-R. Huntzinger and Y.~V.~Kopelevich are gratefully acknowledged. Work supported by PICS CNRS-CNR 2014-2016 Transport phenomena and Proximity-induced Superconductivity in Graphene junctions, Swedish Foundation for Strategic Research (SSF) under the project "Graphene based high frequency electronics", FIRB HybridNanoDev RBFR1236VV (Italy) and by EU FP7, under grant agreement no 604391 Graphene Flagship.

\section{Appendix: Classical simulation of vortex dynamics}

We simulate  random walk of vortices with the  thermal  Brownian dynamics of  $N=2048$ disks  interacting  in 2D space  via  the potential 
\beq
u(r)=\epsilon\left[\left(\frac{\sigma}{r}\right)^2+e^{-r^2/\ell^2}\right].
\label{poi}
\eneq
Here $\epsilon$ is the unit of energy, and $\ell=8\,\sigma$ ( $\sigma $ is the length unit, practically corresponding to the disk diameter).  The disks move in  a 2D square box  of side $L=\sqrt{N/\rho}$, where $\rho$ is the density of the disks. Periodic boundary conditions have been used. The unit of time is  $t_0=\sigma\sqrt{M/\epsilon}$ where $M$ is the mass of the disks. The  free particle diffusivity is $D_0=\sigma^2/t_0$.  After thermalization,  at a temperature $T=0.012\,\epsilon$, we extract the diffusion coefficient $D$ of the interacting disks.
Due to the Einstein relation, the diffusivity $D \propto k_BT / \eta d $ is inversely proportional to the viscosity $\eta$. Here  $d$ is the thickness of the weak link layer.  In Fig.\ \ref{fig:disks} we plot the product $\rho Dt_0$  as a function of $\rho\sigma^{2}$.
We find a low density regime in which the long range Gaussian interaction is dominant  and the diffusion coefficient increases with the density.  
 At higher densities, on the other hand, the short range repulsion becomes dominant, and the diffusion coefficient decreases roughly linearly with the density, giving rise to a constant product $\rho Dt_0$.
 
 To mimic the vortex dynamics, we assume that $\rho \sigma ^2$ of  the disks is proportional to  $H \xi^2 /\phi_0$. Here the coherence length $\xi$  plays the role of  the radius of the vortex core. The corresponding horizontal scale in the plot of   Fig. \ref{fig:disks}  shows  that the crossover  occurs at  $ H_a \sim  0.5$ Oe
 if the  vortex core $\xi \sim 1.3 \: \mu$m.  This points to quite extended Pearl vortices  which are of no surprise in our structure.   Next,
we trade  $\hbar / t_0$  for $k_BT $ in the Einstein relation so that :
\beq
\rho \: D \: t_0 \to \frac{ \hbar}{\eta d } \frac{H}{\phi_0}  = \frac{ (2e)^2}{\hbar} \: R ,
\eneq
where $R = \rho _f /d $ and  Eq. (\ref{vivi}) has been used  in the last equality  to connect the viscosity $\eta $ to  the flux flow resistivity $\rho_f$.
 

\end{document}